\documentclass{aastex}
\usepackage{spr-astr-addons}
\usepackage{url}

\begin{document}
\title{Line shifts in accretion disks - the case of Fe K$\alpha$}
\shorttitle{Line shifts in accretion disks}
\shortauthors{Jovanovi\'{c} et al.}

\author{P. Jovanovi\'{c}} 
\email{pjovanovic@aob.rs}
\affil{Astronomical Observatory, Volgina 7, P.O. Box 74 11060 Belgrade, Serbia}
\and 
\author{V. Borka Jovanovi\'{c}}
\email{vborka@vinca.rs}
\and
\author{D. Borka}
\email{dusborka@vinca.rs@vinca.rs}
\affil{Atomic Physics Laboratory (040), Vin\v{c}a Institute of Nuclear 
Sciences, University of Belgrade, P.O. Box 522, 11001 Belgrade, Serbia}
\and
\author{L. \v C. Popovi{\'c}}
\email{lpopovic@aob.rs}
\affil{Astronomical Observatory, Volgina 7, P.O. Box 74 11060 Belgrade, Serbia}

\begin{abstract}
Here we present a short overview and main results of our investigations of 
several effects which can induce shifts in the broad Fe K$\alpha$ line emitted 
from relativistic accretion disks around single and binary supermassive black 
holes. We used numerical simulations based on ray-tracing method in the Kerr 
metric to study the role of classical Doppler shift, special relativistic 
transverse Doppler shift and Doppler beaming, general relativistic gravitational 
redshift, and perturbations of the disk emissivity in the formation of the 
observed Fe K$\alpha$ line profiles. Besides, we also investigated whether the 
observed line profiles from the binary systems of supermassive black holes could 
be affected by the Doppler shifts due to dynamics of such systems. 
The presented results demonstrate that all these effects could have a 
significant influence on the observed profiles of the broad Fe K$\alpha$ line 
emitted from relativistic accretion disks around single and binary supermassive 
black holes.
\end{abstract}


\section{Introduction}
\label{s:intro}

It is now widely accepted that most galaxies harbor a supermassive black hole 
(SMBH) in their centers and that the formation and evolution of galaxies is 
fundamentally influenced by properties of their central SMBHs 
\citep[see e.g.][and references therein]{kor13}. Active galaxies 
most likely represent one phase in galactic evolution, and they derive their 
extraordinary luminosities from energy release by matter accreting towards, and 
falling into, a central SMBH through a relativistic accretion disk which 
represents an efficient mechanism for extracting the gravitational potential 
energy and converting it into the electromagnetic radiation \citep[see 
e.g.][and references therein]{jov09,jov12}

Since its discovery in Seyfert 1 galaxy MCG-6-30-15 by \citet{tan95}, the broad 
Fe K$\alpha$ emission line at 6.4 keV was observed in a number of Active 
Galactic Nuclei (AGNs) \citep[see e.g.][and reference therein]{nan07}. 
Based on the observations with the X-ray satellites like Ginga, 
ASCA, RXTE, BeppoSAX, XMM-Newton, Suzaku and Chandra, the fact that X-ray 
irradiation of the accretion disk surface in a class of AGNs known as Seyfert 1 
galaxies gives rise to the fluorescent Fe K$\alpha$ line emission via the X-ray 
reflection is now well established \citep[see e.g.][]{fab00}. The broad and 
asymmetric profile of this line with narrow bright ''blue'' and wide faint 
''red'' peak is most commonly attributed to the relativistic effects due to a 
very fast rotation of the emitting material in the innermost regions of the 
accretion disk \citep[see e.g.][and references therein]{fab00}, although there 
are some alternative attempts to explain such profile by 
scattering/reflection/absorption by the disk wind and/or ionized outflows which 
could mimic relativistic effects \cite[see][and references therein]{tur09}. 
Since this line is emitted from a very compact region in the accretion disk 
near the central SMBH, it represents a powerful diagnostic tool for studying 
the effects of strong gravitational field \citep[see e.g.][]{jov08a}, accretion 
physics and space-time geometry in the vicinity of SMBHs (see \citet{jov12} for 
an overview and \citet{jov11} for a case study).

So far, the extensive theoretical studies of the line profiles emitted 
from relativistic accretion disks around SMBHs have been carried out, in which 
the line profiles have been modeled using different approaches \citep[see 
e.g.][for a review]{rey03}. In the weak field limit such line profiles can 
be evaluated analytically \citep{chen89a,chen89b}. \citet{fab89} were the first 
who calculated the Fe K$\alpha$ profiles emitted by the accretion disk in the 
Schwarzschild metric. A full general relativistic approach based on ''the 
transfer-function'' which contains all relativistic effects is also used for 
modeling relativistically broadened lines \citep{laor91}. The exact line 
profiles can be also computed by the direct integration of the photon 
trajectories in the Kerr metric \citep{kar92,bro97} which enabled studies of 
the complex disk models, such as geometrically thick or warped accretion disks 
\citep{har00} or disks with non-axisymmetric structures in form of spiral arms 
\citep{kar01}. The standard X-ray spectral fitting code, XSPEC \citep{arn96} is 
often used for evaluating the general relativistic models of accretion disks 
and simulating the corresponding line profiles \citep[see e.g.][]{dov04}.

Here we study several phenomena which can induce line shifts, and thus cause 
the variability of the profile of the Fe K$\alpha$ line emitted from a 
relativistic accretion disk around a SMBH, using a code based on
ray-tracing method in the Kerr metric \citep[see e.g.][and references 
therein]{cad98,jov12}. This paper is organized as follows: 
in Section \ref{s:lines} we briefly describe the method used to obtain the 
simulated line profiles, in Section \ref{s:shifts} we describe several types of 
line shifts and study different effects which can induce them, and finally in 
Section \ref{s:conc} we point out our conclusions.

\begin{figure}[ht]
\centering
\includegraphics[width=\columnwidth]{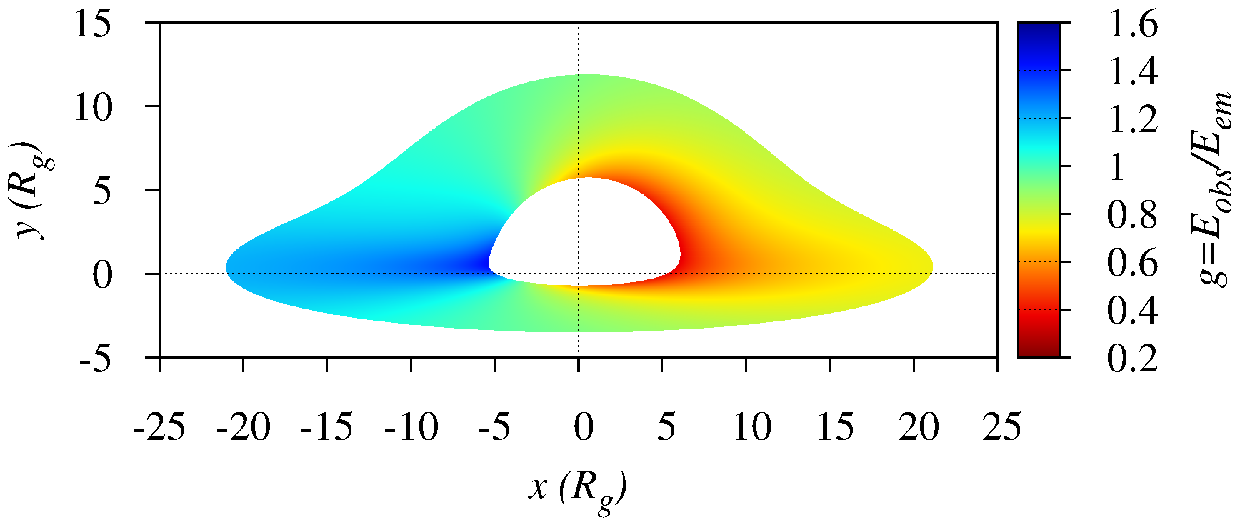} \\
\includegraphics[width=\columnwidth]{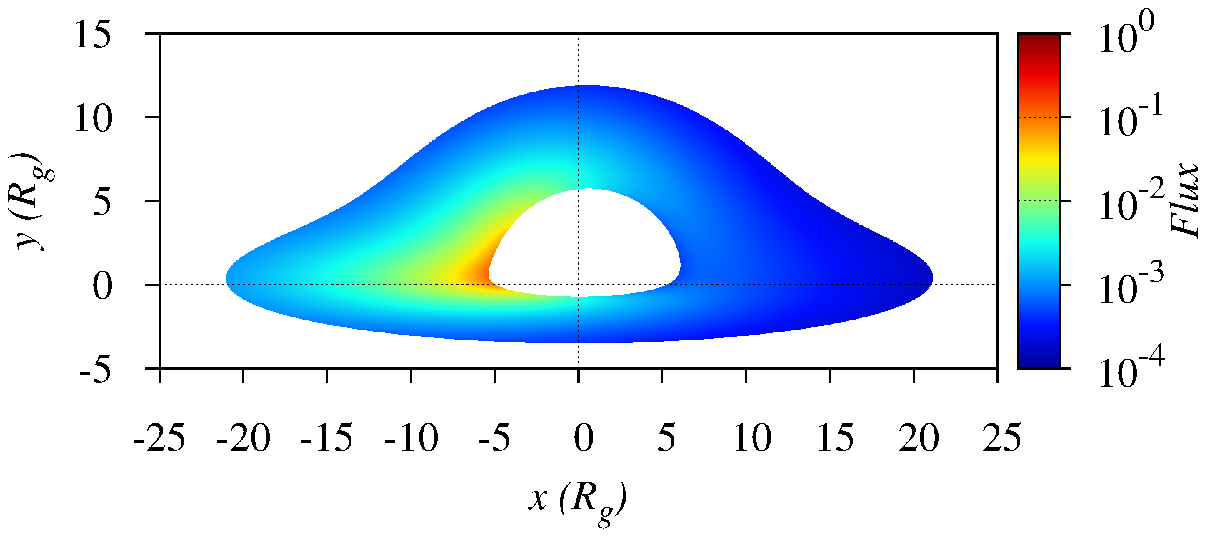}
\caption{Simulated image of a highly inclined ($i=80^\circ$) relativistic 
accretion disk in Kerr metric around a SMBH with spin $a=0.5$, colored 
according to energy shift $g$ (top) and observed flux (bottom). Inner and outer 
radii of the disk are $R_{in}=R_{ms}$ and $R_{out}=20\ R_g$, respectively and 
its power law emissivity index is $q=-2.5$}
\label{fig:disk}
\end{figure}

\begin{table*}[ht]
\centering
\caption{Parameters of the simulated accretion disks, and the intervals 
of energy shift $(g_{min} < g < g_{max})$ over which the corresponding 
simulated line profiles presented in the Figs. \ref{fig:out}-\ref{fig:in} have 
nonzero intensity, as well as the widths of each line bin $\Delta g$. In all 
cases it is assumed that the disk emissivity follows a power law with 
emissivity index $q=2.5$.}
\begin{tabular}{|c|c|c|c|c|c|c|}
\hline
$R_{in}\ (R_g)$ &  $R_{out}\ (R_g)$ &  $i\ (^\circ)$ &  $a$ & 
$g_{min}$  & $g_{max}$ & $\Delta g$ \\
\hline
\hline
7000 & 8000 & 10 & 0.99 & 0.9985 & 1.0012 & 0.0001 \\
\hline
1000 & 2000 & 10 & 0.99 & 0.9935 & 1.0045 & 0.0001 \\
\hline
10 & 20 & 35 & 0.01 & 0.72 & 1.10 & 0.01\\
\hline
6 & 20 & 35 & 0.01 & 0.60 & 1.10  & 0.01 \\
\hline
10 & 20 & 35 & 1 & 0.73 & 1.10 & 0.01 \\
\hline
1 & 20 & 35 & 1 & 0.02 & 1.10 & 0.01 \\
\hline
\end{tabular}
\label{tab:tab1}
\end{table*}

\section{Simulated line shapes}
\label{s:lines}

We studied the radiation from relativistic accretion disks around SMBHs using 
numerical simulations based on ray-tracing method in the Kerr metric 
\citep[see][for more details]{fan97,cad98}, in which we used 
the pseudo-analytical integration of the geodesic equations describing the 
photon trajectories in 
the general case of a rotating SMBH with some angular momentum (spin) $a$.
Due to several effects which will be discussed below, photons emitted from 
a disk at energy $E_{em}$ (or wavelength $\lambda_{em}$) will be observed at 
energy $E_{obs}$ (or wavelength $\lambda_{obs}$) by an observer at infinity, 
causing the energy shift $g$ or, equivalently, the usual redshift in wavelength 
$z$:
\begin{equation}
\label{eqn:shift}
g=\dfrac{E_{obs}}{E_{em}}=\dfrac{\lambda_{em}}{\lambda_{obs}}=\dfrac{1}{1+z}.
\end{equation}
In the ray-tracing method one takes into account only those photon trajectories 
reaching the observer's sky plane. An observer's sky plane is divided into a 
number of small elements (pixels), and for each pixel the photon trajectories 
are traced backward from the observer by following the geodesics in a Kerr 
space-time, until their intersection with the disk. We assumed  
an optically thick and geometrically thin disk for which the spectrum of emitted
radiation  depends on its structure, and therefore on the distance to the 
central SMBH \citep{sha73}. Thus, assuming that disk radiates according to an
emissivity law $\varepsilon \left( {r} \right)$, for each pixel at observer's 
sky plane it is possible to calculate the flux density and the energy shift $g$ 
of the photons emitted by the disk \citep[see][for more details]{jov12}. 
Usually, a power law for disk emissivity is assumed: 
$\varepsilon\left(r\right)\propto r^{-q}$, where $q$ is an emissivity index. In 
that way, one can obtain a simulated image of a relativistic accretion 
disk as would be seen by a distant observer using a powerful high resolution 
telescope. An example of such simulated image of a relativistic accretion disk 
is presented in Fig. \ref{fig:disk}, where  $R_g=GM/c^2$ is the gravitational 
radius of the central SMBH with mass $M$ ($G$ being gravitational 
constant and $c$ speed of light), and $R_{ms}$ represents the radius of the 
marginally stable orbit which depends on spin $a$ of the SMBH.

The corresponding simulated profile of the line emitted from the simulated 
accretion disk can be calculated by taking into account the total 
observed flux and observed photon energies of all pixels over the disk image:
\begin{equation}
\label{eqn:fobs}
F_{obs} \left( {E_{obs}}  \right) = {\int\limits_{image} {\varepsilon
\left({r} \right)}} g^{4}\delta \left( {E_{obs} - gE_{0}}  \right)d\Xi ,
\end{equation}
where $d\Xi$ is the solid angle subtended by the disk in the observer's
sky and $E_{0}$ is the rest energy. Several examples for the simulated line 
profiles emitted from different regions of the relativistic accretion disks with 
different parameters are given in Figs. \ref{fig:out}-\ref{fig:in}.

\section{Results: Fe K$\alpha$ line shifts}
\label{s:shifts}

In this section we will discuss several phenomena which can induce shifts in 
the lines emitted from relativistic accretion disks around SMBHs. 
For that purpose we simulated the line profiles emitted from different 
regions of accretion disks, which parameters are given in Table 
\ref{tab:tab1}. In the following paragraphs we describe the obtained results 
in more details. However, one should take into account that some other 
phenomena, such as gravitational microlensing could also induce line shifts 
\citep[see e.g.][]{pop03a,pop03b,pop06,jov08b,jov09}, but they will not be 
discussed here.

\subsection{Classical Doppler shift due to Keplerian rotation in accretion disk}
\label{ss:kepler}

\begin{figure}[ht]
\centering
\includegraphics[width=\columnwidth]{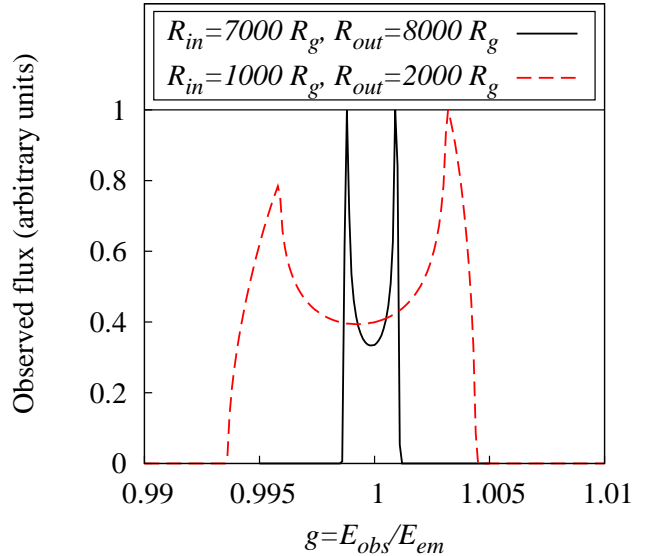}
\caption{Two shapes of the lines emitted from the outer regions of a 
slightly inclined ($i=10^\circ$) relativistic 
accretion disk around a maximally rotating Kerr SMBH, extending between 7000 
and 
8000 $R_g$ (solid line) and 1000 and 2000 $R_g$ (dashed line)}
\label{fig:out}
\end{figure}

Rotation of emitting material in the outermost regions of an accretion disk, 
located thousands $R_g$ from the central SMBH, is practically Keplerian since 
the relativistic effects are negligible at such large distances. Therefore, the 
lines emitted from these regions are broadened depending on Keplerian 
rotational velocity and have two symmetric peaks (see the solid line profile in 
Fig. \ref{fig:out}): the ''blue'' one emitted by the material located on the 
side of the disk which is approaching towards an observer (denoted by blue 
shades in the simulated accretion disk image presented in the top panel of Fig. 
\ref{fig:disk}), and the ''red'' one, emitted by the material on the receding 
side of the disk with respect to the observer (denoted by red shades in the top 
panel of Fig. \ref{fig:disk}).

\subsection{Transverse Doppler shift and Doppler beaming due to special 
relativistic effects}
\label{ss:sr}

However, in the regions which are located less than a few thousand $R_g$ from
the SMBH, the rotational velocities are much higher (on the order of several 
thousand km s$^{-1}$) inducing, not only the wider line shapes, but also the 
special relativistic effects which produce an asymmetry in the line shapes by 
shifting the emission from the inner regions of the disk to the ''red'' and 
beaming its radiation in the direction of motion \citep[see e.g.][]{fab99}, and 
thus enhancing the ''blue'' peak with respect to the ''red'' one (see the 
dashed line profile in Fig. \ref{fig:out}). These two special relativistic 
effects are known as transverse Doppler shift and Doppler beaming, respectively.

The transverse Doppler shift originates when the observer is displaced in a 
direction perpendicular to the direction of the motion of emitters in 
the accretion 
disk, and is a consequence of the relativistic Doppler effect which includes the 
time dilation. Namely, if in the observer's frame the angle between the 
direction of the emitter at the time of emission and the observed direction of 
the light at the time of observation is equal to $\pi/2$, the relativistic 
Doppler effect reduces to the transverse Doppler shift, due to which the 
observed radiation from a rapidly rotating accretion disk is redshifted by the 
Lorentz factor, as it is the case with the dashed simulated line profile in 
Fig. \ref{fig:out}.

Doppler beaming (boosting) modifies the apparent brightness of the rapidly 
moving emitting matter in the sense that, if it is moving towards the observer 
then it will appear brighter than if it is at rest, and vice versa, if it is 
moving away from the observer it will appear fainter than if it is at rest. 
When this is applied to the rapidly rotating emitters in the accretion disk, it 
will enhance the apparent brightness of the approaching side of the disk with 
respect to its receding side (see the bottom panel in Fig. \ref{fig:disk}), and 
thus it will also enhance the ''blue'' peak of the line with respect to its 
''red'' peak, as it can be seen from the dashed simulated line profile in Fig. 
\ref{fig:out}.

Since the velocities of material, as well as the radii of the emitting region 
of the dashed simulated line profile in Fig. \ref{fig:out}, correspond to the 
parts of the disk from which the optical spectral lines like $H\alpha$ and 
$H\beta$ are observed, one can conclude that the transverse Doppler shift and  
Doppler beaming significantly affect even the optical lines which originate 
from outer regions of the accretion disk. In the case of such lines,  
as it can be seen from Table \ref{tab:tab1}, the maximum redshift of the photons
$(1/g_{min}-1)$ is significantly larger by absolute value than their maximum 
blueshift $(1/g_{max}-1)$, which quantitatively represents the influence of 
transverse Doppler shift in this case.  

\begin{figure}[t]
\centering
\includegraphics[width=\columnwidth]{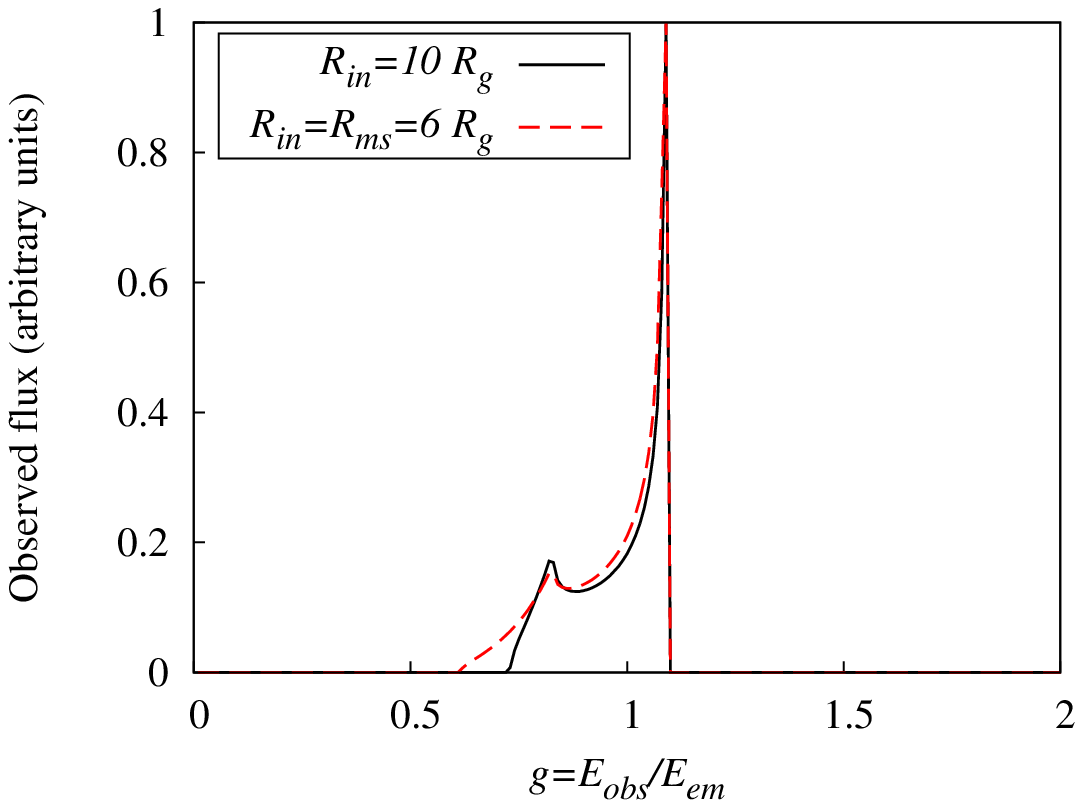} \\
\includegraphics[width=\columnwidth]{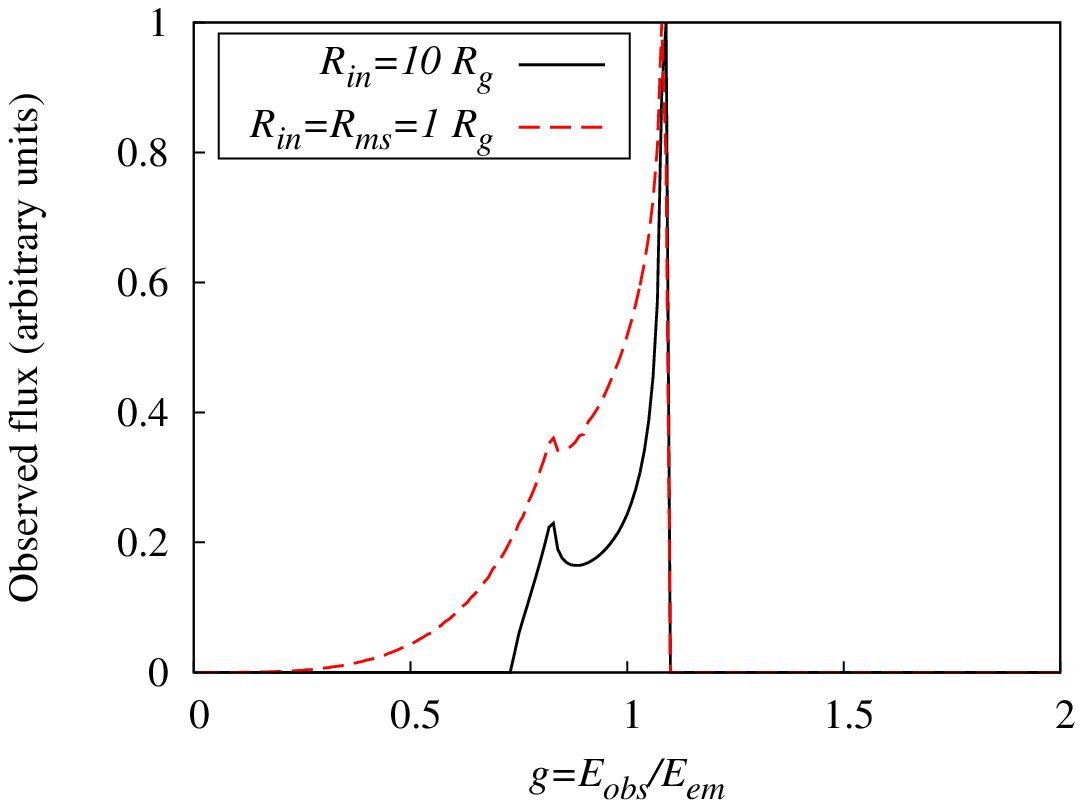}
\caption{\textit{Top:} the shape of the Fe K$\alpha$ line emitted from the 
innermost regions of a relativistic accretion disk around a Schwarzschild SMBH 
with inclination of $i=35^\circ$, outer radius of $20\ 
R_g$ and inner radius of $R_{in}=10\ R_g$ (solid line), as well as 
$R_{in}=R_{ms}=6\ R_g$ (dashed line). \textit{Bottom:} the same as above, but in 
the case of an accretion disk around a maximally rotating Kerr SMBH for which 
$R_{ms}=1\ R_g$}
\label{fig:in}
\end{figure}

However, since the Fe K$\alpha$ line originates in the innermost regions of an 
accretion disk, its emitting material rotates with the fastest velocities (on 
the order of tens of thousands of km s$^{-1}$). Therefore, the previously 
mentioned classical and special relativistic effects are even more significant 
in the case of Fe K$\alpha$ line (see the both panels in Fig. \ref{fig:in}, 
as well as the corresponding rows in Table \ref{tab:tab1}). As 
it can be seen from Fig. \ref{fig:in}, when compared to the simulated profile 
of an optical line (dashed profile from Fig. \ref{fig:out}), the Fe K$\alpha$ 
line profile is much wider (due to the larger classical Doppler shift), more 
asymmetric (due to the larger Doppler beaming) and more redshifted (due to the 
larger transverse Doppler shift, as well as due to the gravitational 
redshift which is discussed in more details in the following paragraph).

\subsection{Gravitational redshift due to general relativistic effects}
\label{ss:gr}

According to the general relativity, the observed frequency of electromagnetic 
radiation emitted by a source in a strong gravitational field in vicinity of a 
SMBH will be reduced with respect to the emitted frequency, due to 
gravitational time dilation. This effect is called the gravitational redshift 
\citep{mis73}, and in the Newtonian limit it is approximately 
proportional to the mass of a SMBH and inversely proportional to the distance 
between it and a point in the disk at which a photon is emitted.

In order to study the influence of the gravitational redshift on the Fe 
K$\alpha$ line profile, we simulated the line profiles originating from 
two emitting regions in accretion disk around a non-rotating Schwarzschild SMBH 
(top panel in Fig. \ref{fig:in}) and a maximally rotating Kerr SMBH (bottom 
panel in Fig. \ref{fig:in}), having two different inner radii: $R_{in}=10\ R_g$ 
(denoted by solid lines) and $R_{in}=R_{ms}$ (denoted by dashed lines). The 
outer radii in both cases are $R_{out}=20\ R_g$. As it can be seen from Fig. 
\ref{fig:in}, in the case of both metrics the profiles originating from 
emitting regions with smaller inner radii are redshifted with respect to the 
profiles from emitting regions with larger inner radii, due to the 
gravitational redshift. This is even more obvious if one compare the 
corresponding maximum redshifts $(1/g_{min}-1)$ and blueshifts $(1/g_{max}-1)$ 
presented in Table \ref{tab:tab1}, from which it can be seen that the 
maximum redshifts are much bigger in the cases with the smaller inner radii, 
while the blueshifts are the same in both cases. 
Therefore, the gravitational redshift causes the further deformations of the Fe 
K$\alpha$ line profile by smearing its ''blue'' emission into the ''red'' one.

This effect is stronger in the case of a maximally rotating Kerr SMBH than in 
the case of a non-rotating Schwarzschild SMBH due to the fact that the radius of 
the innermost stable circular orbit (also known as radius of marginally stable 
orbit, $R_{ms}$) strongly depends on SMBH spin $a$, so that in the case of a 
non-rotating Schwarzschild SMBH (i.e. for $a=0$) it is equal to $6\ R_g$ (top 
panel in Fig. \ref{fig:in}), and in the case of an extremely rotating Kerr SMBH 
(i.e. for $a=1$) it is equal to $1\ R_g$ (bottom panel in Fig. \ref{fig:in}). 
Thus, in the case of an extremely rotating SMBH it is possible to detect the 
radiation which originates closer to the SMBH itself, and which is therefore 
more subjected to the gravitational redshift (see the last row in 
Table \ref{tab:tab1}), than in the case of a non-rotating Schwarzschild SMBH.

In that way the SMBH spin $a$ affects the radiation originating from the 
innermost regions of the disk which are closest to the SMBH \citep[see 
e.g.][]{jov08a} and, as it can be seen from both panels in Fig. \ref{fig:in} 
and the corresponding rows of Table \ref{tab:tab1}, 
it significantly affects the profile of the Fe K$\alpha$ line, especially its 
''red'' wing by extending it towards the lower energies for the higher values 
of $a$ (see also \citet{jov11} for a case study and \citet{jov12} for more 
details).

\subsection{Line shifts due to perturbed disk emissivity}
\label{ss:pert}

Beside the previously mentioned classical, special and general 
relativistic effects, several observational/theo\-re\-ti\-cal studies indicated
that the line shapes and shifts could be affected by the perturbations of the 
disk emissivity in the form of flares or orbiting bright spots, especially in 
the case of high accretion rate objects, such as Seyfert galaxies and QSOs 
\citep[see e.g.][]{cze04a,cze04b,jov10}. For instance, appearance of the 
redshifted Fe K$\alpha$ emission feature due to co-rotating flare above the 
accretion disk which irradiates a spot on its surface was reported in the case 
of the Seyfert galaxies NGC 3516 \citep{iwa04} and NGC 3783 \citep{tom07}, 
based on their observations by XMM-Newton. However, since the sensitivity of 
XMM-Newton was not sufficient to clearly resolve sub-orbital features in the 
observed X-ray spectra, the azimuthal irradiation structure of the inner 
accretion disk is modeled and simulated by \citet{cze04a,cze04b,gos07a,gos07b} 
in order to investigate the detectability of orbiting spots in nearby Seyfert 
galaxies with current and future X-ray observatories, such as X-ray Evolving 
Universe Spectroscopy (XEUS), a planned successor to XMM-Newton which evolved to 
Advanced Telescope for High Energy Astrophysics (ATHENA+) that is currently 
under development, aiming to have several hundred times larger sensitivity than 
Chandra X-ray Observatory and XMM-Newton. For that purpose, 
\citet{gos07a,gos07b} developed a method to compute the local spectra of the 
hot spot emission from the surface of an accretion disk underlying a strong 
co-rotating flare above the disk, and used it to model the X-ray reprocessing 
from a persisting flare lasting for a significant fraction of one orbital period 
at the given disk radius in some AGNs such as MCG-6-30-15 and NGC 5548, as well 
as the short-term flares with durations of a few hundreds of seconds. These 
studies showed that the future high precision observations will allow to 
disentangle the effects of the flares in the accretion disks of Seyfert galaxies 
from the intrinsic variability of the local emission, even at distances which 
are less than $5\ R_g$ from the central SMBHs \citep{gos07b}. Also, the spectral 
variations induced by such flares depend on the viewing direction and are 
stronger for intermediate (edge-on) viewing angles, as supposed for Seyfert-2 
galaxies, than for a face-on viewing direction, as assumed in Seyfert-1 
galaxies, and thus, they could be used to put observational constraints on the 
disk inclination \citep{cze04a,cze04b}. However, indications for the flares in 
the accretion disks were found not only in the X-ray band, but also in the 
optical spectra of some AGNs. For example, the variability of the observed 
$H\beta$ line profiles of quasar 3C 390.3 was recently successfully explained by 
a perturbation of the power law disk emissivity, revealing the appearance of two 
successive bright spots on the approaching side of the disk which were 
interpreted as the fragments of spiral arms in the accretion disk of 3C 390.3 
\citep{jov10}.

\begin{figure}[ht]
\centering
\includegraphics[width=\columnwidth]{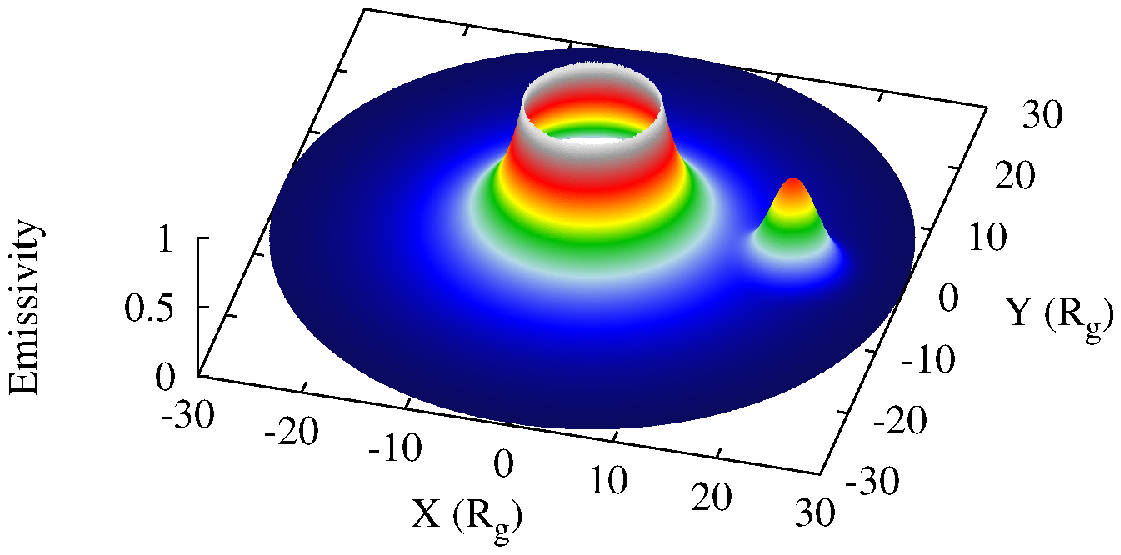} \\
\includegraphics[width=\columnwidth]{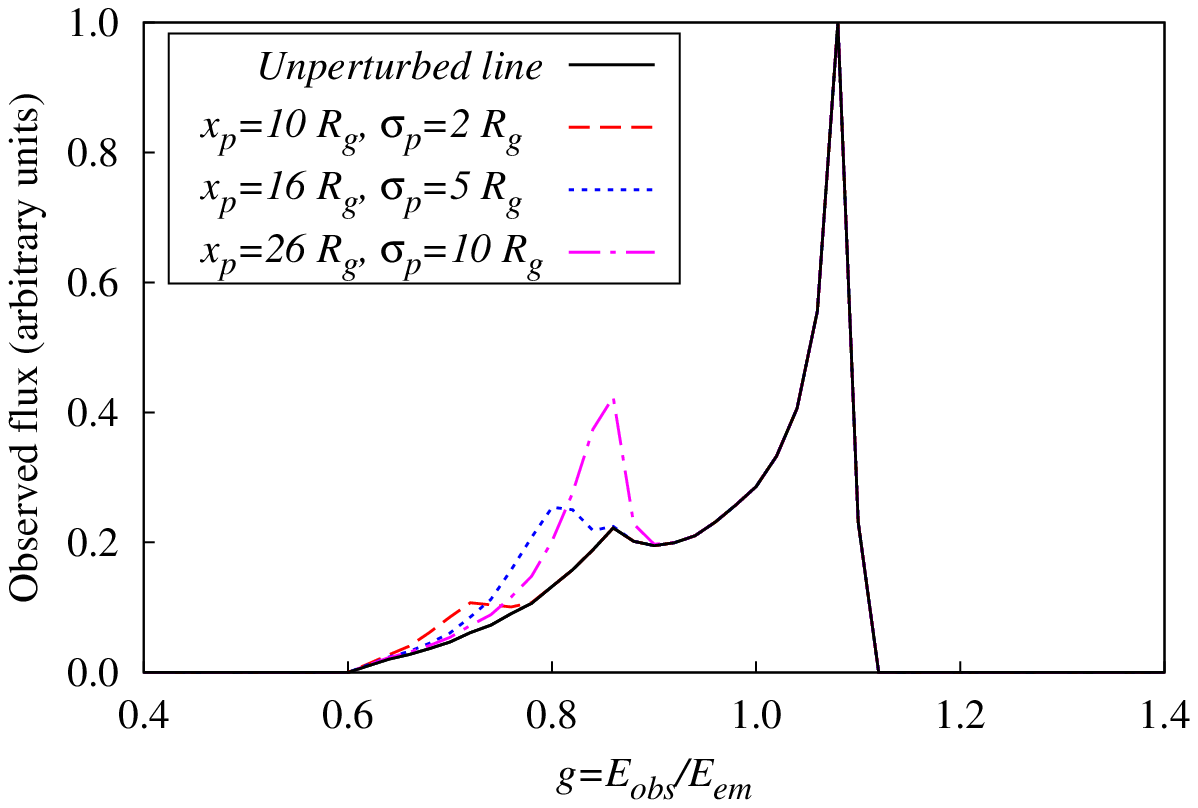}
\caption{\textit{Top:} illustration of perturbed emissivity of an accretion
disk in Schwarzschild metric with inclination of 
$i=35^\circ$, extending from $R_{ms}$ up to $30\ R_g$ for a 
perturbing region located at the receding side of the disk. \textit{Bottom:} 
the unperturbed (solid line) and perturbed (dashed lines) shapes of the Fe 
K$\alpha$ line for a perturbing region located at three different positions 
along $x$-axis and with three different widths: a) $x_p=10\ R_g$, $\sigma_p=2\ 
R_g$, b) $x_p=16\ R_g$, $\sigma_p=5\ R_g$ and c) $x_p=26\ R_g$, $\sigma_p=10\ 
R_g$}
\label{fig:pert}
\end{figure}

Such perturbations of the power law disk emissivity can be modeled as 
\citep{jov10}:
\begin{equation}
\label{eqn:pert}
\varepsilon\left(x,y\right)\propto
r\left(x,y\right)^{-q}\cdot\left(1+\varepsilon_p\cdot 
e^{-\frac{(x-x_p)^2+(y-y_p)^2}{\sigma_p^2}}\right),
\end{equation}
where $q$ is power law emissivity index, $\varepsilon_p$ is emissivity of 
perturbing region, and $(x_p,y_p)$ and $\sigma_p$ are its position and width 
(both in $R_g$), respectively.
An illustration of the perturbed disk emissivity is presented in the top panel 
of Fig. \ref{fig:pert}, and the perturbed and unperturbed Fe K$\alpha$ line 
profiles for three different positions and widths of perturbing region are given 
in the bottom panel of the same figure. As one can see from Fig. \ref{fig:pert}, 
a perturbing region on the receding side of accretion disk, affects only the 
''red wing'' of the Fe K$\alpha$ line, while the ''blue'' one and the line core 
stay nearly unaffected. Moreover, the position of perturbing region is in a
direct relation with the position of the ''red peak'' of the line. 
Recently, a model of accretion disk with perturbing region was used for 
fitting the observed spectra of some AGNs, resulting with a non-random 
distribution of bright spot positions along the receding side of the disk. This 
indicated that the most plausible physical mechanism which can explain such 
behaviour is a fragmented spiral arm of the accretion disk \citep[see e.g.][and 
references therein]{jov10,lew10}. Perturbing regions could be then explained by 
the emissivity lumps caused by fragments in spiral arms, such as isolated 
clumps which could pass through the arm and dominate in its emissivity. The 
fragments naturally travel with the arm as it precesses through the disk, but 
since the outer parts of an accretion disk are self-gravitating, the fragments 
of spiral arms could be also launched from the outer regions of the disk and 
propagate inward \citep{lew10}. Besides, in the inner regions of the disk, the 
disk winds and highly ionized fast outflows could play an important role and 
could also cause the outward propagation of such clumps \citep[see 
e.g.][]{pop11,sim10}. Since the flux variability caused by a spiral arm is on 
timescales of a year to several years while its fragments cause shorter 
variations on timescales of several months, this model is able to explain the 
observed variability of some AGNs on the timescales ranging from several months 
to several years \citep{lew10}. As it can be seen from Figure \ref{fig:pert}, 
when a perturbation which originates in the inner regions of the disk and, due 
to some wind or outflow, spirals away along the receding side of the disk 
further from the central SMBH and toward the outer parts of the disk, it will 
cause the blueward shift of the ''red peak'' which will therefore move towards 
the line core. This demonstrates that in some cases, the deformations of the 
line profiles could arise due to shifts of some of their parts, caused by the 
variability of the disk emissivity \citep[for more details see][]{jov10}.

\subsection{Line shifts in the binary systems of SMBHs}
\label{ss:smbb}

Binary systems of supermassive black holes are formed in galactic mergers, and
at some stage when two SMBHs become gravitationally bound and start to orbit 
around their center of mass, accretion of the surrounding matter on both SMBHs 
could be expected and as a result, a strong line emission could arise (see 
\citet{pop12} for theoretical aspects and \citet{bon12,yan15} for observational 
evidences). In such a case, the Fe K$\alpha$ line might be also observed, and 
it would most likely arise from both accretion disks around primary and 
secondary SMBHs, located in a central low density cavity of a circumbinary disk 
\citep{jov14}. Therefore, the Fe K$\alpha$ line profiles emitted from both 
accretion disks might be affected by the Doppler shifts due to the orbital 
motion of the SMBH binary, and in that case the corresponding signatures would 
be imprinted in the observed composite line profile \citep{jov14}.

\begin{figure}[ht]
\centering
\includegraphics[width=\columnwidth]{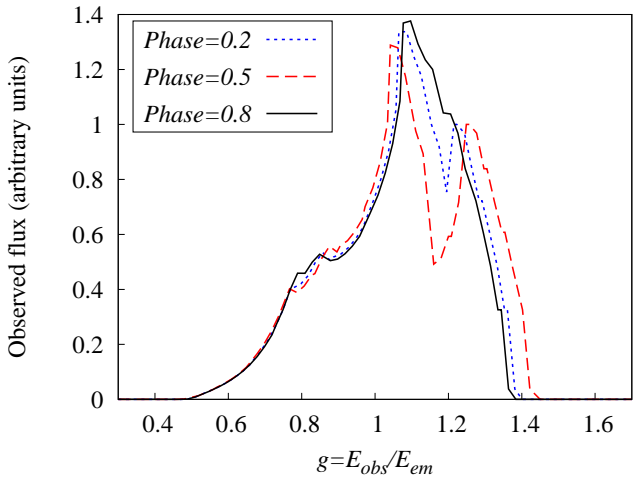}
\caption{Composite profiles of the Fe K$\alpha$ line emitted from the accretion 
disks around two SMBHs in a binary system during three different phases of its 
Keplerian orbit. The parameters of the SMBH binary system are: mass ratio 
$q=1$, separation between the components of $a=0.01$ pc, inclination 
$i=60^\circ$, eccentricity $e=0.75$, longitude of pericenter $\omega=90^\circ$ 
and systemic velocity $\gamma=0$. Both accretion disks surround the slowly 
rotating Kerr primary and secondary SMBHs with the same spins $J/(M c)=0.1$ (M 
being the mass of a SMBH), and both of them extend from $R_{ms}$ to $30\ R_g$, 
have the same inclinations of $60^\circ$, as well as power law emissivity 
indices of $p=2$}
\label{fig:smbb}
\end{figure}

Assuming that the radial velocities of the primary and secondary components 
are $V_{1,2}^{rad} \ll c$ so that the corresponding Doppler shifts due to 
their orbital motion are $\approx V_{1,2}^{rad} / c$, a composite profile 
$F\left(g\right)$ of the Fe K$\alpha$ line emitted from both accretion disks of 
a SMBH binary at some orbital phase can be calculated from two constituent 
unshifted line profiles $F_1 \left(g\right)$ and $F_2 \left(g\right)$ according 
to \citep{jov14}:
\begin{equation}
\label{eqn:smbb}
F\left(g\right)=F_1\left(\left[\frac{1}{g}-\frac{V_1^{rad}}{c}\right]^{-1}
\right)+F_2\left(\left[\frac{1}{g}-\frac{V_2^{rad}}{c}\right]^{-1}\right).
\end{equation}

Three such simulated composite Fe K$\alpha$ line profiles emitted during 
three different phases along the Keplerian orbit of a SMBH binary with 
sub-parsec separation between the components are presented in Fig. 
\ref{fig:smbb}. As it can be seen from Fig. \ref{fig:smbb}, Doppler shifts due 
to the orbital motion of the components in the binary have significant 
influence on the observed composite Fe K$\alpha$ line profiles and induce their 
ripple variability, as well as the shifts of their parts, especially of the 
line cores and their ''blue'' wings \citep[for more details see][]{jov14}. Such 
shifts and ripple variability of the complex Fe K$\alpha$ line profiles 
strongly depend on the orbital phase of the binary system, and therefore if 
observed in the spectra of some AGNs, they could represent the observational 
signatures of the existence of SMBH binary systems in the cores of these AGNs.

\section{Conclusions}
\label{s:conc}

Here we discussed different types of the Fe K$\alpha$ line shifts in 
AGNs, as well as several phenomena that can induce them. For that 
purpose, we performed numerical simulations of the X-ray radiation from the 
relativistic accretion disks around single and binary SMBHs, based on 
ray-tracing method in the Kerr metric. The obtained results showed that the 
following effects have a significant influence on the line profiles:
\begin{itemize}
\item 
classical Doppler shift which causes double-peaked profiles of the lines 
emitted over a whole relativistic accretion disk, from its innermost 
regions where X-ray radiation originates, to its outermost parts which 
emit in the optical band;
\item 
special relativistic transverse Doppler shift and relativistic beaming which 
cause the redshift of the line profile, as well as the relative enhancement of 
its ''blue'' peak with respect to the ''red'' one, not only in the case of the 
broad Fe K$\alpha$ line which originates from the innermost regions of the disk, 
but also in the case of the optical lines which originate from its outer parts;
\item 
general relativistic gravitational redshift which causes the further 
deformations of the Fe K$\alpha$ line profile by smearing its ''blue'' emission 
into the ''red'' one, as well as the SMBH spin $a$ which affects the ''red'' 
wing of the line by extending it towards the lower energies for the higher 
values of $a$;
\item
perturbations of the disk emissivity in form of flares or bright spots moving 
along the receding side of the disk, which can induce blueward 
shift of the ''red peak'' not only of the Fe K$\alpha$ line, but also of the 
lines in other spectral bands;  
\item
Doppler shifts due to the orbital motion in the SMBH binaries which could 
produce the ripple variability of the complex Fe K$\alpha$ line profiles and 
induce shifting of their parts. Such variability of the observed Fe K$\alpha$ 
line profiles strongly depends on the orbital phases of the SMBH binary 
systems, and could represent their observational signatures.
\end{itemize}

\acknowledgments
This research is part of the projects 176003 ''Gravitation and the large scale 
structure of the Universe'' and 176001 ''Astrophysical Spectroscopy of 
Extragalactic Objects'', supported by the Ministry of Education, Science and 
Technological Development of the Republic of Serbia. This research is also 
partially supported by the SEENET-MTP - ICTP Program PRJ-09 ''Strings and 
Cosmology''.

\end{document}